\documentstyle[12pt,twoside,fleqn,espcrc1,epsfig]{article}

\newcommand{\AmS}{{\protect\the\textfont2
  A\kern-.1667em\lower.5ex\hbox{M}\kern-.125emS}}

\title{Hyperon polarization in Kaon Photoproduction from the deuteron.}

\author{H.~Yamamura${}^{\rm a}$, K.~Miyagawa
\address{Department of Applied Physics, Okayama University of Science, 1-1 Ridai-cho, Okayama 700, Japan}%
, T.~Mart
\address{Jurusan, Fisika, FMIPA, Universitas Indonesia, Depok 16424, Indonesia}%
, C.~Bennhold ${}^{\rm c}$, H. Haberzettl
\address{Center for Nuclear Studies, Department of Physcs, The George Washington University, Washington, D.C. 20052}%
, W.~Gl\"ockle
\address{Institut f\"ur Theoretische Physik II, Ruhr-Universit\"at Bochum, D-44780 Bochum, Germany}}

\begin{document}
% typeset front matter
\maketitle

\begin{abstract}
We analyze the reaction $\gamma + d \rightarrow K^+ + \Lambda (\Sigma) + N$ using the Nijmegen soft core hyperon-nucleon interactions NSC97f and NSC89 and a recently updated production operator for the $\gamma + N \rightarrow K^+ + \Lambda(\Sigma)$ processes. Significant effects of the $YN$ final state interaction are found near both the $K^+ \Lambda N$ and $K^+ \Sigma N$ thresholds.
\end{abstract}

\section{Introduction}

Since hyperon-nucleon scattering experiments are difficult to perform, hyperon production processes such as $\gamma + d \rightarrow K^+ + Y + N$ and $e + d \rightarrow e^\prime + K^+ + Y + N$ appear as natural candidates that allow exploring the $YN$ interaction. One can obtain the information of the $YN$ interaction by analyzing the correlated $YN$ final states. An inclusive $d(e,e^\prime K^+ )YN$ experiment has already been performed in Hall C at TJNAF, while the data for $d(\gamma, K^+ Y)N$ are being analyzed in Hall B.

Recently, we found that various meson-theoretical $YN$ interactions generate $S$-matrix poles around the $\Lambda N$ and $\Sigma N$ thresholds\cite{miya2}. The pole near the $\Sigma N$ threshold is related to the strength and the property of the $\Lambda N - \Sigma N$ coupling and causes enhancements in the $\Lambda N$ elastic total cross sections. The hope is that the pole structure of the $YN$ $t$ matrix will have visible effects in such production processes mention above.

In this paper, we study the inclusive $d(\gamma, K^+)YN$ and exclusive $d(\gamma, K^+ Y)N$ processes for $\theta_K=0^\circ$ and predict various observables including polarization observables.

\section{Formalism}

The reaction processes $\gamma + d \rightarrow K^+ + \Lambda (\Sigma) + N$ are expressed by the operator $T_i$ as
\begin{equation}
T_i | \Psi_d > = \sum_j U_{ij} \cdot t_{\gamma K}^{(j)} | \Psi_d >\hspace{10mm} i,j=\Lambda N,\Sigma N,
\label{tmat1}
\end{equation}
where the operator $t_{\gamma K}^{(i)}$ describes the elementary processes $\gamma + N \rightarrow K^+ + \Lambda (\Sigma)$, and $|\Psi_d>$ represents the deuteron state which is generated by the Nijmegen 93 $NN$ interaction \cite{nsc93nn}. The operator $U_{ij}$ corresponds to the $YN$ final state interaction processes, and is represented as
\begin{eqnarray}
U_{ij}&=&\delta_{ij} + V_{ij}G_0^{(j)} + \sum_{j^\prime} V_{ij^\prime} G_0^{(j^\prime)} V_{j^\prime j} G_0^{(j)} + \cdots \nonumber \\
&=& \delta_{ij} + \sum_{j^\prime} V_{ij^\prime} G_0^{(j^\prime)} U_{j^\prime j},
\label{uij}
\end{eqnarray}
where $V_{ij}$ is $YN$ interaction including $\Lambda N - \Sigma N$ coupling. We ignore the $K^+$ meson interaction with the nucleon and hyperon in the final states. From Eqs.(\ref{tmat1}) and (\ref{uij}), one can deduce the coupled set of integral equations for $T_i$,
\begin{equation}
T_i |\Psi_d> = t_{\gamma K}^{(i)} |\Psi_d> + \sum_{j^\prime} V_{ij^\prime} G_0^{(j^\prime)} T_{j^\prime} |\Psi_d>.
\label{tmat2}
\end{equation}
We solve this set (\ref{tmat2}) after partial-wave decomposition in momentum space. The three elementary process $\gamma + p \rightarrow K^+ + \Lambda (\Sigma^0)$ and $\gamma + n \rightarrow K^+ + \Sigma^-$ are properly incorporated in the driving term in Eq.(\ref{tmat2}). Equation (\ref{tmat2}) is solved on isospin bases $\Lambda N$ and $\Sigma N$, but the resulting amplitudes are transformed into those on the particle bases $\Lambda n$, $\Sigma^0 n$ and $\Sigma^- p$ by which the inclusive $d(\gamma,K^+)$, exclusive $d(\gamma,K^+Y)$ cross sections and hyperon polarizations are calculated. For details, we refer the reader to ref.\cite{yama1}.

\section{Results}

\begin{figure}[t]
\begin{center}
\epsfig{file=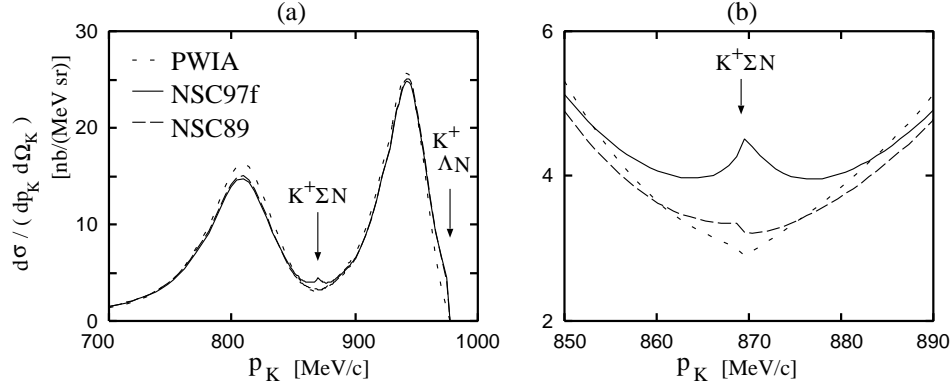,width=125mm}
\end{center}
\caption{Inclusive $d(\gamma ,K^+)$ cross section as a function of lab momentum $p_{K^+}$ for $\theta_{\rm K^+,lab}=0^\circ$ and the photon lab energy $E_\gamma=1.3$ GeV. The two thresholds $K^+ \Lambda N$ and $K^+ \Sigma N$ are indicated by the arrows. The results around the $K^+ \Sigma N$ threshold are enlarged in (b).}
\label{fsi1}
\end{figure}

At present, we calculate the observables only for the $K^+$ meson scattered to 0 degree $(\theta_{K^+}=0^\circ)$. The Nijmegen soft-core $YN$ interactions NSC97f\cite{nsc97} and NSC89\cite{nsc89} and a recently updated production operator\cite{elba} for the $\gamma+N\rightarrow K^+ + \Lambda(\Sigma) + N$ processes are used. 

Figure \ref{fsi1}(a) shows the inclusive cross sections which sum up the contributions of the $K^+ \Lambda n$, $K^+ \Sigma^0 n$ and $K^+ \Sigma^- p$ final states. The solid and dashed lines are the predictions of the NSC97f and NSC89 $YN$ interactions, respectively. The dotted line shows the results of the plane wave impulse approximation (PWIA). The arrows indicate the two thresholds $K^+ \Lambda N$ ($p_K=977.30$ MeV/c) and $K^+ \Sigma N$ ($p_K=869.14$ MeV/c).

The two pronounced peaks around $p_K=$945 and 809 MeV/c are due to the quasi-free processes of $\Lambda$ and $\Sigma$, where one of the nucleon in the deuteron is spectator and has zero momentum in the laboratory system. Significant FSI effects are found around the $K^+ \Lambda N$ and $K^+ \Sigma N$ thresholds. The cross section is increased up to 86\% by FSI near the $K^+ \Lambda N$ threshold. Around the $K^+ \Sigma N$ threshold, as shown Fig.\ref{fsi1}(b), the strengths and shapes of the enhancements by the NSC97f and NSC89 are quite different.

\begin{figure}[htb]
\begin{center}
\epsfig{file=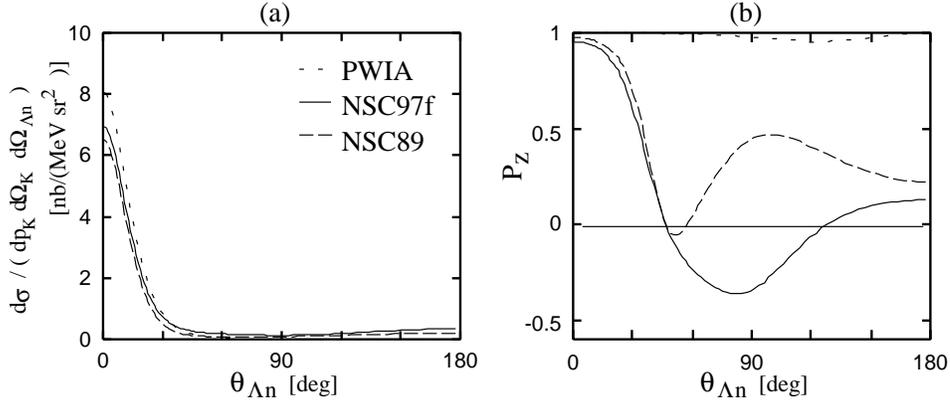,width=125mm}
\end{center}
\caption{(a) Exclusive $d(\gamma ,K^+ \Lambda)$ cross section and (b) $\Lambda$ recoil polarization with incoming polarized photon as a function of the $\Lambda$ scattering angle in the $\Lambda n$ c.m. system. The photon lab energy is $E_\gamma = 1.3$ GeV. The outgoing kaon lab momentum and angle are $p_{K^+}=870$ MeV/c and $\theta_{K^+}=0^\circ$ respectively.}
\label{expol1}
\end{figure}

\begin{figure}[htb]
\begin{center}
\epsfig{file=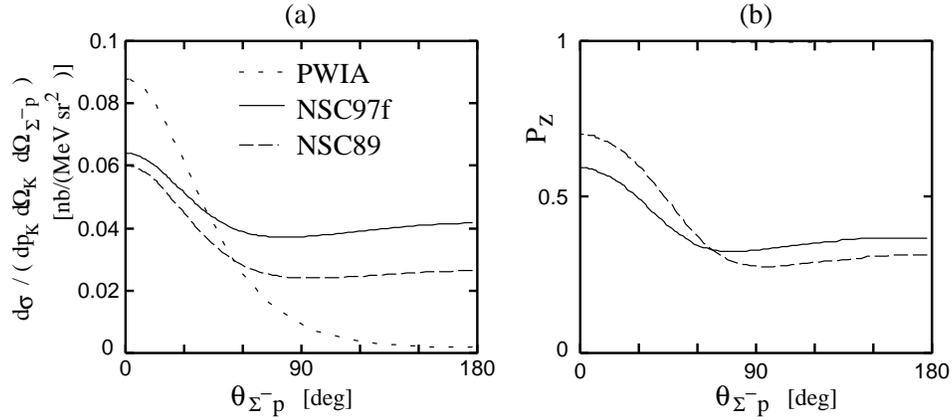,width=125mm}
\end{center}
\caption{(a) Exclusive $d(\gamma ,K^+ \Sigma^-)$ cross section and (b) $\Sigma^-$ recoil polarization with incoming polarized photon as a function of the $\Sigma^-$ scattering angle in the $\Sigma^- p$ c.m. system. The photon lab energy is $E_\gamma =1.3$ GeV. The outgoing kaon lab momentum and angle are $P_{K^+}=865$ MeV/c and $\theta_{K^+}=0^\circ$ respectively.}
\label{expol2}
\end{figure}

\begin{figure}[htb]
\begin{center}
\epsfig{file=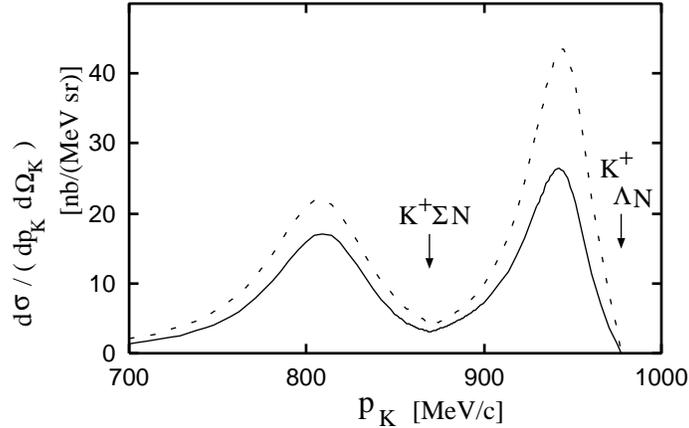,width=90mm}
\end{center}
\caption{Inclusive $d(\gamma, K^+)$ cross section in the PWIA as a function of lab momentum $p_{K^+}$ for $\theta_{K^+}=0^\circ$ and photon lab energy $E_\gamma=1.3$ GeV. The solid curve shows the prediction by the present version\cite{elba} of the production operator for $\gamma + N \rightarrow K^+ + \Lambda(\Sigma)$, while the dashed line corresponds to the prediction by an old version\cite{nphysic}.}
\label{newold}
\end{figure}

Figure \ref{expol1}(a) illustrates the exclusive $d(\gamma, K^+ \Lambda)$ cross sections just below the $K^+ \Sigma N$ threshold ($p_K=870$ MeV/c). The FSI effects are seen both at very forward and at large angles. The PWIA cross sections are basically zero at backward angles, while the FSI calculations still show some strength. Figure \ref{expol1}(b) demonstrates the $\Lambda$ recoil polarizations with incoming polarized photon. The $\Lambda$ recoil polarizations in PWIA are almost one. This is because the incoming photon is polarized along the $z$ axis, but the target deuteron is unpolarized and the outgoing $K^+$ meson carries no spin and angular momentum in this case ($\theta_K=0^\circ$). However, the final state interactions cause the large deviations from one, and the prediction by NSC97f is quite different from that of NSC89. 

\clearpage

The exclusive results to the $K^+ \Sigma^- p$ final states just above this threshold ($p_K=$865 MeV/c) are shown in Fig.\ref{expol2}. The prominent FSI effects are seen both in the cross sections and in the double polarization observable.

Finally, we briefly discuss the production operator for $\gamma + N \rightarrow K^+ + \Lambda(\Sigma)$ processes. In Fig.\ref{newold}, the inclusive cross sections in PWIA in which an old version\cite{nphysic} of the operator is used are compared to those with the present version\cite{elba}. The latter has been improved in the fitting to the data of $\gamma + N \rightarrow K^+ + \Lambda(\Sigma)$ including the new SAPHIR data. The difference between the predictions by the two versions are quite large as in Fig.\ref{newold}, which suggests this reaction $d(\gamma,K^+)YN$ is another promising candidate for investigating the operator.


\begin{thebibliography}{99}
\bibitem{miya2} K.~Miyagawa, H.~Yamamura, Phys. Rev. C60 (1999) 024003; nucl-th/9904002.
\bibitem{nsc93nn}   V. G. J. Stoks, R. A. M. Klomp, C. P. F. Terheggen, and J. J. de~Swart, Phys. Rev. C49 (1994) 2950.
\bibitem{yama1}H.~Yamamura, K.~Miyagawa, T.~Mart, C.~Bennhold, W.~Gl\"ockle, Phys. Rev. C61 (1999) 014001; nucl-th/9907029.
\bibitem{nsc97}   Th. A. Rijken, V. G. J. Stoks, and Y.~Yamamoto,
   Phys. Rev. C59 (1999) 21.
\bibitem{nsc89}   P. M. M.~Maessen, Th. A.~Rijken, and J. J.~de~Swart,
   Phys. Rev. C40 (1989) 2226.
\bibitem{elba}
   C. Bennhold, T. Mart, A. Waluyo, H. Haberzettl, G. Penner,
   T. Feuster, and U. Mosel, in {\it Proceedings of the Workshop
   on Electron-Nucleus Scattering, Elba, Italy, 1998}, 
   edited by O. Benhar, A. Fabrocini, and R. Schiavilla
   (Edizioni ETS, Pisa, 1999), p. 149; nucl-th/9901066.
\bibitem{nphysic}
   C.~Bennhold, T.~Mart, and D.~Kusno, in {\it Proceedings of the
   CEBAF/INT Workshop on $N^*$ Physics, Seattle, USA, 1996},
   edited by T.-S.~H.~Lee and W.~Roberts (World Scientific, Singapore,1997), p.166
\end{thebibliography}
\end{document}